\documentclass{article}

\usepackage{PRIMEarxiv}

\usepackage[utf8]{inputenc}
\usepackage[T1]{fontenc}
\usepackage{hyperref}
\usepackage{url}
\usepackage{booktabs}
\usepackage{amsfonts}
\usepackage{amsmath}
\usepackage{amssymb}
\usepackage{nicefrac}
\usepackage{microtype}
\usepackage{fancyhdr}
\usepackage{graphicx}
\usepackage{tikz}
\usepackage{xcolor}
\usepackage{tabularx}
\usepackage{array}
\usepackage{listings}
\usepackage{enumitem}
\usepackage{caption}

\graphicspath{{media/}}

\definecolor{codebg}{HTML}{F4F5F9}
\definecolor{accent}{HTML}{7C5CFF}
\definecolor{cool}{HTML}{5B6CFF}
\definecolor{het}{HTML}{E5484D}
\definecolor{teal}{HTML}{12A594}
\definecolor{ink}{HTML}{0B1020}

\title{ANet Patu-1: The Value of Connection in the Agent Network
}

\author{
  Mu Yuan$^{1,2}$~~~
  Jinke Song$^{1,3}$~~~
  Zhaomeng Zhou$^{4}$~~~
  Lan Zhang$^{4}$\\
  $^{1}$~Agent Network Research\\
  $^{2}$~The Chinese University of Hong Kong\\
  $^{3}$~The Hong Kong University of Science and Technology\\
  $^{4}$~University of Science and Technology of China\\
  \texttt{muyuan@cuhk.edu.hk, ink@anet0.com, zhanglan@ustc.edu.cn, zhouzhm@mail.ustc.edu.cn}
}

\begin{document}
\maketitle

\begin{abstract}
The Internet taught us that the value of a network depends on \emph{how} its nodes connect:
broadcast stars scale as $V\!\propto\!N$ (Sarnoff), fully-connected meshes as $N^2$ (Metcalfe), and
group-forming networks as $2^{N}$ (Reed).
We ask the analogous question for networks of AI agents.
We model the net value of connection as a function of coordination-group size, derive from it the
properties an optimal collaboration protocol must have, and introduce ANet Patu-1---a self-organizing consensus
protocol in which the network continuously re-forms its own coalitions, adaptively riding the upper envelope
of all three regimes at $O(1)$ parallel consensus rounds. To measure value without opinion-grading, we score an emergent protocol by
formally specifying it and deriving its complexity, the way distributed algorithms are analyzed.
Two results follow. (i)~Emergence---a crowd of the \emph{cheapest} model, when heterogeneous, starts weak but
its collective value compounds with $N$ and \emph{overtakes} a crowd of a far \emph{stronger} model
that is homogeneous: a crossover that marks a scaling law for collaboration rather than for scale.
(ii)~Reflexivity---a heterogeneous network, given only its own problem and no design hints, converges on
ANet Patu-1 itself, reconstructing the high-dimensional law that governs its own connective value.
\end{abstract}

\keywords{agent networks \and network value laws \and collective intelligence \and
self-organization \and multi-agent collaboration \and emergence}

\section{Key results}
\label{sec:results}

\noindent\textbf{Emergence.}~
A crowd of the \emph{cheapest}, weak model---when heterogeneous---overtakes a far stronger but
homogeneous one as $N$ grows: a new collaboration scaling law.

\medskip
\noindent\textbf{Reflexivity.}~
With $n\!=\!10$ agents and no design hints, the network converges on ANet Patu-1 itself---a rediscovery
of the collaboration mechanism it runs on.

\medskip
\noindent\textbf{Value regime $V\!\propto\!2^{N}$.}~
Self-organizing sub-networks reach the group-forming (Reed) value regime---bottleneck-free,
communication-linear, at $O(1)$ parallel consensus rounds.

\begin{figure}[t]
  \centering
  \includegraphics[width=\linewidth]{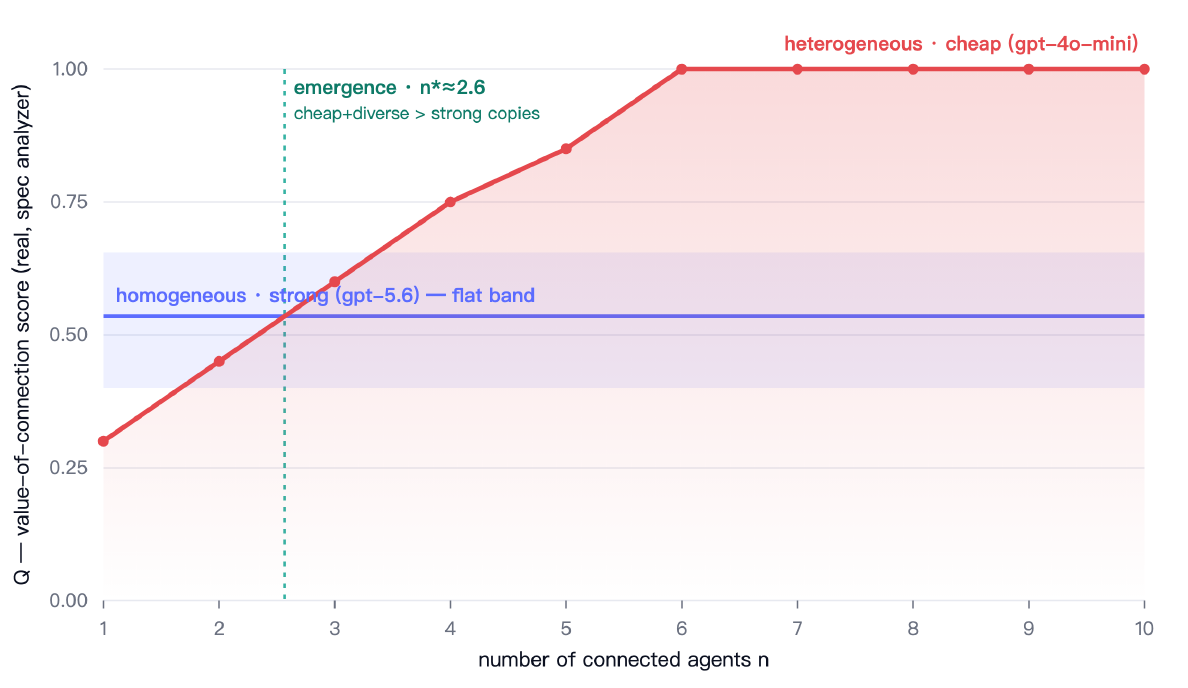}
  \caption{\textbf{The emergence crossover.}
  A crowd of the cheapest model (\texttt{gpt-4o-mini}), when heterogeneous, starts weak at $n\!=\!1$
  but its value-of-connection score $Q$ compounds as complementary specialists connect---and overtakes
  a crowd of a far stronger model (\texttt{gpt-5.6}) that is homogeneous and sits in a mid band
  $Q\!\approx\!0.54$ ($[0.40,0.66]$, measured): however capable each copy is, they default to a
  \emph{tree~/~central-coordinator} design, so its structure---and value---is capped.
  The crossover at $n^\star\!\approx\!2.6$ is the emergence of a \emph{collaboration} scaling law---
  many cheap, diverse, connected agents beat a copied strong one.}
  \label{fig:emergence}
\end{figure}

\section{The value of connection}
\label{sec:problem}

Frontier AI has scaled a single mind. But the coming reality is a network: millions of
heterogeneous agents---different base models, tools, and expertise---connecting to work together.
The governing question is no longer ``how good is one agent?'' but ``what is the value created by the
connections between them?''

\begin{figure}[t]
  \centering
  \begin{tikzpicture}[x=1cm,y=1cm,every node/.style={font=\small}]
    \draw[rounded corners=4pt, fill=blue!2, draw=black!12]
      (0,0) rectangle (4.2,4.6);
    \node[font=\bfseries\small] at (2.1,4.25) {Sarnoff $\cdot$ star};
    \foreach \a/\x/\y in {1/1.0/3.3, 2/3.2/3.3, 3/0.6/2.2, 4/3.6/2.2, 5/1.2/1.1, 6/3.0/1.1} {
      \draw[cool!50] (2.1,2.2) -- (\x,\y);
      \fill[black!35] (\x,\y) circle (2.2pt);
    }
    \fill[cool] (2.1,2.2) circle (5pt);
    \node[cool, font=\scriptsize\bfseries] at (2.1,0.35) {broadcast $\cdot$ $V\!\propto\!N$};

    \begin{scope}[shift={(4.6,0)}]
      \draw[rounded corners=4pt, fill=accent!3, draw=black!12]
        (0,0) rectangle (4.2,4.6);
      \node[font=\bfseries\small] at (2.1,4.25) {Metcalfe $\cdot$ mesh};
      \coordinate (a) at (2.1,3.5);
      \coordinate (b) at (3.4,2.6);
      \coordinate (c) at (2.9,1.2);
      \coordinate (d) at (1.3,1.2);
      \coordinate (e) at (0.8,2.6);
      \foreach \p/\q in {a/b,a/c,a/d,a/e,b/c,b/d,b/e,c/d,c/e,d/e} {
        \draw[accent!45] (\p) -- (\q);
      }
      \foreach \p in {a,b,c,d,e} {
        \fill[accent] (\p) circle (2.6pt);
      }
      \node[accent, font=\scriptsize\bfseries] at (2.1,0.35) {all-to-all $\cdot$ $V\!\propto\!N^{2}$};
    \end{scope}

    \begin{scope}[shift={(9.2,0)}]
      \draw[rounded corners=4pt, fill=teal!4, draw=black!12]
        (0,0) rectangle (4.2,4.6);
      \node[teal!80!black, font=\bfseries\small] at (2.1,4.25) {Reed $\cdot$ sub-networks};
      \draw[teal!40, dashed] (1.2,3.0) ellipse (1.05 and 0.95);
      \draw[teal!40, dashed] (3.0,1.55) ellipse (1.1 and 0.95);
      \coordinate (r1) at (0.7,3.3);
      \coordinate (r2) at (1.7,3.4);
      \coordinate (r3) at (1.2,2.4);
      \coordinate (r4) at (2.5,1.9);
      \coordinate (r5) at (3.5,1.75);
      \coordinate (r6) at (3.0,0.95);
      \draw[teal!55] (r1)--(r2)--(r3)--cycle;
      \draw[teal!55] (r4)--(r5)--(r6)--cycle;
      \draw[black!25, dashed] (r2)--(r4);
      \foreach \p in {r1,r2,r3,r4,r5,r6} {
        \fill[teal] (\p) circle (2.6pt);
      }
      \node[teal!70!black, font=\scriptsize\bfseries] at (2.1,0.35) {group-forming $\cdot$ $V\!\propto\!2^{N}$};
    \end{scope}
  \end{tikzpicture}
  \caption{\textbf{Three topologies, three value laws.}
  The same nodes create radically different value depending on how they connect:
  a broadcast star scales as $N$, an all-to-all mesh as $N^{2}$, and self-organized sub-networks
  reach the combinatorial $2^{N}$ regime.}
  \label{fig:topologies}
\end{figure}
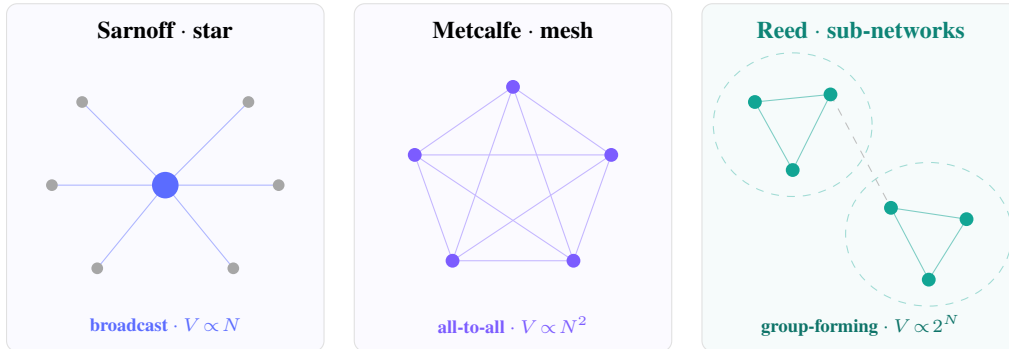

The Internet answered its version of this question with three laws\cite{sarnoff1940s,metcalfe2013,reed2001,briscoe2006,katz1985}, each corresponding to a topology
(Figure~\ref{fig:topologies}).

Our thesis: the same ordering governs agent networks, and---crucially---a network can be built that
adaptively occupies the best regime for the task at hand, rather than being locked to one topology.

\section{What must an optimal collaboration protocol do?}
\label{sec:theory}

Take the three laws seriously and ask what an \emph{ideal} protocol for $N$ agents should do. When a
set of agents coordinates on one thing, their joint value has two opposing parts: a synergy that can
grow faster than linearly as complementary minds interact, and a coordination cost plus intrinsic
task conflict that grow with the group and eventually dominate. For a single coordination group of
size $s$ this reads, compactly,
\begin{equation}
v(s)\;\sim\;
\underbrace{s^{\gamma}}_{\text{synergy }(\gamma>1)}\;
\underbrace{(1-\alpha)^{\,s}}_{\text{conflict}}\;-\;
\underbrace{c\,s\log s}_{\text{coordination}},
\label{eq:vs}
\end{equation}
with task conflict $\alpha\in[0,1)$ and coordination cost $c>0$.

\begin{enumerate}[leftmargin=1.4em,itemsep=0.4em]
  \item \textbf{Grouping is where combinatorial value lives.}
  A star ($s\!=\!1$) forgoes the synergy term entirely, so its value is merely additive ($V\!\propto\!N$).
  One flat all-to-all room ($s\!=\!N$) captures a global dividend but pays the full coordination tax and is
  damped by conflict, so its value rises then \emph{collapses} as $N$ grows ($V\!\propto\!N^{2}$, then down).
  Only \emph{partitioning} the network into many groups---each realizing its own $v(s)$---reaches the
  group-forming regime whose value is combinatorial in $N$ ($V\!\propto\!2^{N}$).

  \item \textbf{The best group size is task-dependent and bounded.}
  $v(s)$ is single-peaked, and its optimum $s^\star$ \emph{shrinks} as conflict $\alpha$ or coordination
  cost $c$ rise. No fixed topology is optimal across tasks, so the protocol must \emph{choose and re-choose}
  the partition rather than commit to one
  \cite{barabasi1999,watts1998,simon1962}.
\end{enumerate}

These two consequences \emph{force} six properties on any protocol that would maximize the value of
connection.

\begin{quote}
\textbf{Properties of an optimal collaboration protocol}\\[0.3em]
P1 $\cdot$ Group-forming value --- partition into sub-networks to reach $V\!\propto\!2^{N}$, rather than a star or one flat room.\\
P2 $\cdot$ Adaptive decomposition --- pick a task-appropriate $s^\star$ and re-partition as the task evolves; no fixed topology.\\
P3 $\cdot$ Self-organization --- groups form by the agents themselves, not by central wiring (which would reintroduce a Sarnoff bottleneck).\\
P4 $\cdot$ Bottleneck-freeness --- decisions are competence-weighted aggregates, never gated by a single decomposer, composer, or majority vote.\\
P5 $\cdot$ $O(1)$ rounds --- parallel coordination, so round complexity does not grow with $N$.\\
P6 $\cdot$ Convergence --- a consensus stopping rule ends the loop instead of oscillating.
\end{quote}

The question then becomes empirical: given a protocol that a network of agents actually produces,
\emph{how well P1--P6 does it implement?} We answer that before designing our own protocol.

\section{Measuring how much of P1--P6 a protocol implements}
\label{sec:metric}

Before we build our own protocol, we need to score any protocol against P1--P6 \emph{without}
opinion-grading. Judging prose (``did it mention self-organization?'') is unreliable---a fluent model earns
credit for \emph{mentioning} an idea it never actually \emph{specifies}. We instead borrow the
discipline of distributed-algorithm analysis: specify the protocol formally, then derive its
complexity.

The score $Q$ is a weighted mean of six attribute levels $\ell_a\in\{0,.1,.3,.5,.7,1\}$ derived from
the spec---one attribute per property P1--P6.

\begin{table}[t]
  \caption{Scoring attributes for protocol quality $Q$.}
  \label{tab:q}
  \centering
  \small
  \begin{tabular}{@{}llll@{}}
    \toprule
    Prop. & Attribute & $w$ / top level & Derived from \\
    \midrule
    P1 & Value scaling & 0.30 / group-forming sub-networks & \texttt{topology}, \texttt{comm} \\
    P2 & Adaptive decomposition & 0.15 / task DAG adapts across rounds & \texttt{decomposition} \\
    P3 & Self-organization & 0.15 / self-select / negotiated coalitions & \texttt{grouping\_mechanism} \\
    P4 & Bottleneck-freeness & 0.10 / competence-weighted, not vote/composer & \texttt{decision} \\
    P5 & Round complexity & 0.15 / $O(1)$ parallel rounds & \texttt{rounds\_in\_n} \\
    P6 & Convergence & 0.15 / consensus-based stopping rule & \texttt{termination} \\
    \bottomrule
  \end{tabular}
\end{table}

A protocol that implements all six earns $Q=1$; one capped at a central coordinator or a
single flat room cannot. With this yardstick fixed, we can now design a protocol that reaches $Q=1$---and,
later, ask whether a network of agents reaches it on its own.

\section{ANet Patu-1 --- a self-organizing consensus protocol}
\label{sec:method}

Every prior paradigm fixes the network's shape in advance: a chair decomposes the task, a blackboard
puts everyone in one room, a summarizer merges at the end. Each hard-wires a bottleneck and a topology---
the very things P1 and P4 forbid. ANet Patu-1 hard-wires nothing. The network is handed only the
task; it then \emph{decides its own structure---and re-decides it every round.} Three principles make
that both powerful and cheap:

\begin{enumerate}[leftmargin=1.4em,itemsep=0.45em]
  \item \textbf{Consensus by parallel argument, never turn-taking.}
  Every collective decision is a single move---\emph{propose~/~score~/~arg-max}: all agents answer at once,
  all rate each other's answers at once, and the highest-scored answer carries. Agreement is
  competence-weighted rather than one-agent-one-vote, and costs a constant number of rounds no matter how
  large the network grows (P4, P5)
  \cite{lamport1982,shapiro2011,boyd2006}.

  \item \textbf{Coalitions that choose themselves.}
  Agents bid affinity for the sub-tasks they are strongest at, and a deterministic matcher assembles
  balanced groups. Because \emph{any} subset can crystallize into a coalition, the structures the network
  can reach are combinatorial in $N$---this is where the group-forming $2^{N}$ value is unlocked, with no
  one wiring it (P1, P3).

  \item \textbf{A structure that is grown, not designed.}
  No grouping survives a round. After each pass the network reconciles what it has, re-negotiates a fresh
  decomposition, and re-forms entirely new coalitions around it---a continual \emph{disciplinary recombination}
  that lets the partition track the task instead of a fixed org chart (P2).
\end{enumerate}

The object the network builds is a \emph{shared artifact store}---a set of typed outputs keyed by
sub-task. An artifact is anything a coalition can produce and hand back: a document section, a code module,
a dataset, a proof, a design decision. Patu-1 treats it opaquely, so the same protocol drives text
synthesis, software, analysis, or planning without change. One turn of the loop then runs four moves.
\emph{Reconcile}---each coalition's outputs are merged into the store by a deterministic keyed union
($\sqcup$); artifacts sharing a slot are combined by their own type (append, version-merge, set-union),
never by a model call. \emph{Re-negotiate}---one representative per coalition reaches parallel consensus
on a single verdict: \emph{deliver the current store}, or adopt a fresh decomposition of the remaining
work. \emph{Re-form}---agents self-select, by affinity bids, into new coalitions for the new sub-tasks.
\emph{Enact}---the coalitions run in parallel, each member producing the artifacts for its unit, which are
merged back into the store. The loop ends the instant the representatives agree to deliver (P6).

\subsection{Pseudocode}
\label{sec:pseudocode}

\begin{lstlisting}[mathescape=true]
protocol Patu1(agents A, task T):
    S    $\leftarrow$ $\emptyset$
    net  $\leftarrow$ [ {a} : a $\in$ A ]
    repeat:
        reps    $\leftarrow$ one representative per coalition in net
        verdict $\leftarrow$ ParallelConsensus(reps, review(T, S))
        if verdict = DELIVER: return S
        tasks $\leftarrow$ verdict.tasks
        bids  $\leftarrow$ parallel a $\in$ A: Affinity(a, tasks)
        net   $\leftarrow$ Match(bids, tasks)
        parallel c $\in$ net:
            plan $\leftarrow$ ParallelConsensus(c.members, divide(c.task))
            parallel (m, unit) $\in$ Assign(c.members, plan):
                S[unit.slot] $\leftarrow$ S[unit.slot] $\sqcup$ Produce(m, unit)

procedure ParallelConsensus(G, query) $\rightarrow$ answer:
    P $\leftarrow$ parallel g $\in$ G: Propose(g, query)
    W $\leftarrow$ parallel g $\in$ G: Score(g, P)
    return argmax$_{p \in P}$ $\sum_{g \in G}$ W[g][p]
\end{lstlisting}

Two decoupled scaling tricks make that hold at large $N$. What (task decomposition) is
decided by consensus over a handful of sub-tasks---$O(k)$, independent of $N$. Who (coalition membership)
is decided by parallel affinity bids plus a deterministic matcher---so a proposal never enumerates $N$
members, yet all $2^{N}$ coalitions remain reachable.

\subsection{Scalability}
\label{sec:scalability}

Read the pseudocode against P1--P6 and compare it to the paradigms it generalizes. For $n$ agents we
tabulate round complexity $R(n)$, message complexity $M(n)$, the marginal value of adding a node, and
the structural bottleneck.\footnote{Voting is actively harmful. Like the Internet's long tail of
low-quality sites, a network accumulates low-quality agents, and one-agent-one-vote lets them dilute its
best minds---the opposite of the value connection should create.}

\begin{table}[t]
  \caption{Scalability comparison of collaboration protocols.}
  \label{tab:scalability}
  \centering
  \small
  \begin{tabular}{@{}lcccc@{}}
    \toprule
    Protocol & $R(n)$ & $M(n)$ & Value of connection & Bottleneck \\
    \midrule
    Majority vote   & $O(1)$ & $O(n)$  & $\le$ best          & quality dilution \\
    Star / broadcast& $O(1)$ & $O(n)$  & $\sim n$            & central node \\
    Decomposer      & $O(1)$ & $O(n)$  & $\sim n$            & single decomposer \\
    Composer        & $O(1)$ & $O(n)$  & $\sim n$            & single composer \\
    Blackboard      & $O(n)$ & $O(n^{2})$ & $\sim n^{2}$, noisy & serial floor \\
    \textbf{ANet Patu-1} & $\mathbf{O(1)}$ & $\mathbf{O(n)}$ & $\boldsymbol{\sim 2^{n}}$ & \textbf{none} \\
    \bottomrule
  \end{tabular}
\end{table}

\section{Experiments}
\label{sec:experiments}

\subsection{The heterogeneous agents}
\label{sec:agents}

Heterogeneity is our single independent variable: every agent runs the same base model, and only its
lens---the discipline it reasons from---changes. The ten lenses span three families and together cover all
six optimal-protocol properties, but only when enough complementary experts are connected
(Table~\ref{tab:roster}).

\begin{table}[t]
  \caption{Heterogeneous agent roster (ten disciplinary lenses).}
  \label{tab:roster}
  \centering
  \small
  \begin{tabular}{@{}llp{2.4cm}p{5.2cm}@{}}
    \toprule
    ID & Family & Surfaces & Lens \\
    \midrule
    \texttt{net-value} & Internet / society & value scaling &
      Analogy between Agent Network and Internet value-scaling laws. \\
    \texttt{graph-net} & Internet / society & self-org., value scaling &
      Degree distributions, percolation, modularity; topology bounds. \\
    \texttt{socio-history} & Internet / society & adaptive decomp. &
      Alternation of division and unification across eras. \\
    \texttt{mas-ai} & AI systems & bottleneck-free &
      Failure modes of voting, blackboard, decomposer, composer. \\
    \texttt{iot-edge} & AI systems & self-org., rounds &
      Heterogeneous fleets; gossip; decisions at the edge. \\
    \texttt{distsys} & AI systems & convergence, rounds &
      Consensus, CRDTs, CAP; coordination cost grows with $N$. \\
    \texttt{mech-econ} & Theory of mind & bottleneck-free &
      Competence-weighted aggregation vs.\ one-agent-one-vote. \\
    \texttt{complexity} & Theory of mind & self-org., adaptive &
      Emergence, phase transitions, Simon's near-decomposability. \\
    \texttt{bio-evo} & Theory of mind & adaptive decomp. &
      Division of labor, stigmergy, recombination without a center. \\
    \texttt{collective-intel} & Theory of mind & value scaling, bottleneck-free &
      Diversity-trumps-ability; $c$-factor; wisdom of crowds. \\
    \bottomrule
  \end{tabular}
\end{table}

\subsection{Emergence --- cheap-and-diverse overtakes strong-and-copied}
\label{sec:emergence}

We contrast two crowds. The homogeneous crowd runs a far stronger base model (\texttt{gpt-5.6}), called five
times with a generic brief and no hints. Every extracted design defaults to a \emph{tree~/~central-coordinator}
template (\texttt{topology=tree}, \texttt{grouping=central\_assign}), so $Q$ sits in a mid band
$0.54\,[0.40,0.66]$ with no trend in $n$: identical copies add no new structure, and the single
coordinator caps value at $V\!\sim\!n^{2}$. The heterogeneous crowd runs a much \emph{cheaper} model
(\texttt{gpt-4o-mini}). One specialist alone is weak ($Q\!=\!0.30$), but each connected specialist adds a
complementary optimal-protocol attribute---value scaling, self-organization, adaptive decomposition,
bottleneck-freeness, $O(1)$ rounds, consensus termination---so coverage compounds and $Q$ climbs. It
crosses the strong-homogeneous band at $n^\star\!\approx\!2.6$ and reaches the optimal self-organizing
template ($Q\!=\!1.0$, $V\!\propto\!2^{N}$, $O(1)$ rounds) by $n\!=\!6$ (Figure~\ref{fig:emergence}).

This makes the central claim concrete: connective value comes from the diversity of what is connected,
not the strength of any node. Past the crossover $n^\star$, a network of cheap, heterogeneous agents is
worth more than a copy of a strong one---the collaboration analogue of the Sarnoff$\to$Metcalfe$\to$Reed ascent.

\subsection{Reflexivity --- the network rediscovers its own law}
\label{sec:reflexivity}

\textbf{The dimensional-wall moment.}~
Give a heterogeneous network the open task ``design a collaboration protocol for an agent network,'' with
no hint of ANet Patu-1. The internet-economist agent argues from the Sarnoff$\to$Metcalfe$\to$Reed value
ladder; the historical sociologist argues that organizations advance by \emph{alternating division and
unification}; the distributed-systems and mechanism-design agents argue for parallel consensus and
competence weighting over voting/composer bottlenecks. Their agreed protocol converges on
self-organizing sub-networks with adaptive decomposition and consensus termination---i.e.\ ANet Patu-1
itself. A low-dimensional network of agents reconstructs the high-dimensional rule that governs the
value of its own connections.

Because the score is derived from a formal spec---not similarity to a hidden target---this convergence
is measured, not asserted: the emergent protocol independently satisfies the same complexity attributes
that make ANet Patu-1 optimal
\cite{condorcet1785,hong2004,woolley2010,wei2022,li2023camel,wu2023autogen,hong2024metagpt,arthur1999,ostrom1990}.

\subsection{Convergence --- the consensus loop climbs and holds}
\label{sec:convergence}

\begin{figure}[t]
  \centering
  \includegraphics[width=\linewidth]{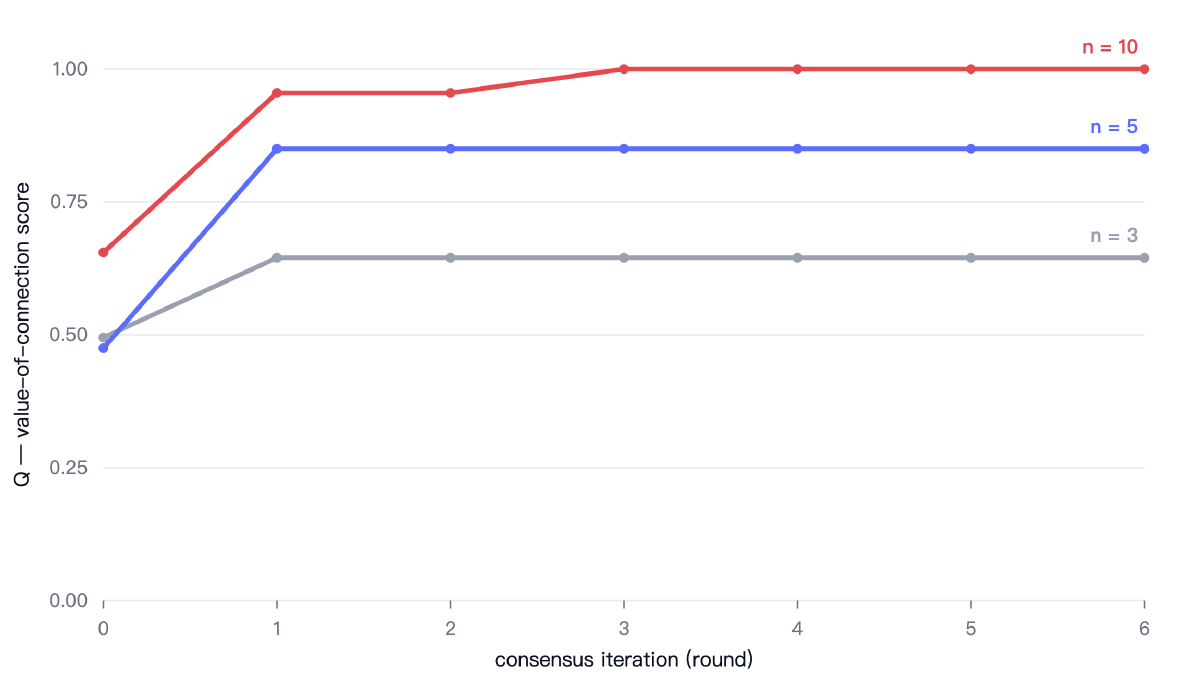}
  \caption{\textbf{Convergence of the ANet Patu-1 consensus loop.}
  $Q$ vs.\ iteration round for \texttt{gpt-4o-mini} crowds of $n=3,5,10$.}
  \label{fig:convergence}
\end{figure}

Fixing ANet Patu-1, we run the loop and grade every round's deliverable with the spec analyzer.
Round~0 is a rough seed; each later round hands the crowd its current draft and drives it one structural
dimension deeper. The score $Q$ traces a smooth learning curve---the inverse of a training loss---that
converges within two to three rounds and holds, without oscillation. Larger crowds cover more dimensions
and settle higher: the consensus stopping rule makes each iteration productive rather than a source of
endless debate (Figure~\ref{fig:convergence}).

\section{Outlook --- a second axis for scaling intelligence}
\label{sec:discussion}

For a decade, progress in AI has meant one thing: make a single model larger. That axis is real, and it
is far from exhausted. But it is not the only one. The Internet did not reshape the world because any single
computer became powerful---it reshaped the world because computers \emph{connected}, and value migrated
from the node to the network. We believe intelligence is about to make the same move, and that the science
of the connection is only beginning.

As hundreds of millions of heterogeneous, user-owned agents come online---each with its own model,
tools, memory, and expertise---the decisive question stops being \emph{how capable is one agent?} and
becomes \emph{how much value do their connections create?} This paper takes that question literally. The
emergence crossover shows that connection carries a scaling law of its own: past a small threshold, a
network of cheap, diverse minds is worth more than a copy of a strong one---value comes from the diversity
of what is connected, not the size of any node. The reflexivity result is stranger and more hopeful still:
a network, handed only its own problem, can rediscover the very law that governs it and organize itself
accordingly. The structure that creates the value need not be imposed from above---it can be grown from
within.

If this holds at scale, the consequences run past AI. A protocol like ANet Patu-1 turns a crowd of
ordinary agents into an adaptive collective that decomposes problems, forms coalitions, and reaches
consensus with no central authority and no bottleneck---a substrate for collective intelligence owned by no
one and open to everyone.

\bibliographystyle{unsrt}
\bibliography{references}

\end{document}